\documentclass[a4paper,11pt]{article}
\usepackage{pos}
\usepackage{amsmath}
\NewDocumentCommand\Nf{mggg}{$N_{\textrm{f}}\text{=}#1\IfNoValueTF{#2}{}{\text{+}#2}\IfNoValueTF{#3}{}{\text{+}#3}\IfNoValueTF{#4}{}{\text{+}#4}$}

\title{Isovector axial and pseudoscalar form factors from twisted mass
  lattice QCD at the physical point}
\ShortTitle{Isovector axial FFs}

\author[a,b]{Constantia Alexandrou}
\author[b]{Simone Bacchio}
\author[c]{Martha Constantinou}
\author[d]{Jacob Finkenrath}
\author[e]{Roberto Frezzotti}
\author[f]{Bartosz Kostrzewa}
\author*[b]{Giannis Koutsou}
\author[a]{Gregoris Spanoudes}
\author[f]{Carsten Urbach}

\affiliation[a]{Department of Physics, University of Cyprus}
\affiliation[b]{Computation-based Science and Technology Research Center, The Cyprus Institute}
\affiliation[c]{Department of Physics, Temple University, Philadelphia}
\affiliation[d]{Department of Theoretical Physics, European Organization for Nuclear Research, CERN}
\affiliation[e]{Dipartimento di Fisica and INFN, Universit\'a di Roma ``Tor Vergata'', Rome}
\affiliation[f]{HISKP (Theory), Rheinische Friedrich-Wilhelms-Universit\"at Bonn}

\emailAdd{g.koutsou@cyi.ac.cy}

\abstract{We present the isovector axial, induced pseudoscalar, and
  pseudoscalar form factors of the nucleon using three twisted-mass
  fermion ensembles with degenerate up- and down-, strange-, and
  charm-quarks with masses tuned to their physical values (physical
  point). The three ensembles have lattice spacing $a$=0.08, 0.068,
  and 0.057~fm and approximately equal physical volume allowing for
  the continuum limit to be taken at the physical point. Excited-state
  contributions to the matrix elements are evaluated using several
  sink-source separations from 0.5~fm to 1.5~fm and multistate
  fits. We check the partially conserved axial-vector current (PCAC)
  hypothesis and the pion pole dominance (PPD) and show that in the
  continuum limit both relations are satisfied. We provide results at
  the continuum limit for the isovector nucleon axial charge, axial
  radius, pion-nucleon coupling constant, and for the induced
  pseudoscalar form factor at the muon capture point.}

\FullConference{The 41st International Symposium on Lattice Field Theory (LATTICE2024)\\
 28 July - 3 August 2024\\
Liverpool, UK\\}

\begin{document}
\maketitle

\section{Introduction}
The nucleon axial form factors are fundamental quantities that
characterize the nucleon's response to weak interactions and play a
crucial role in neutrino scattering experiments. They are particularly
relevant for current and upcoming neutrino experiments such as
NO$\nu$A, MINER$\nu$A, and DUNE. At Fermi Lab, the MINER$\nu$A
experiment has recently provided new measurements of neutrino
interactions~\cite{MINERvA:2023avz}. While the axial charge $g_A$ is
well determined from neutron beta decay
experiments~\cite{Brown:2017mhw,Darius:2017arh,Mendenhall:2012tz,Mund:2012fq},
the momentum dependence of the axial form factor $G_A(Q^2)$ and the
induced pseudoscalar form factor $G_P(Q^2)$ are less well constrained
by experiment.

Lattice QCD provides a first-principles approach to computing these
quantities directly from the QCD Lagrangian. Recent progress has
enabled simulations at physical quark masses, eliminating the need for
chiral extrapolation which can introduce uncontrolled systematic
uncertainties. Early lattice studies were limited to the quenched
approximation~\cite{Liu:1991nk,Liu:1992ab} or heavier than physical
pion masses~\cite{Alexandrou:2007eyf}.  In this work, we present the
first calculation to use solely simulations at physical pion mass to
take the continuum limit, using three ensembles generated with twisted
mass fermions~\cite{Alexandrou:2023qbg}. The ensembles span lattice
spacings from 0.08~fm to 0.057~fm, enabling a controlled continuum
extrapolation of all quantities. We perform a thorough analysis of
excited-state contributions and examine important relations such as
the partially conserved axial-vector current (PCAC) and pion pole
dominance (PPD).

\section{Axial and pseudo-scalar form factors}

The nucleon matrix element of the isovector axial-vector current
$A_\mu = \bar{u} \gamma_\mu \gamma_5 u -\bar{d}\gamma_\mu \gamma_5 d$
can be decomposed in terms of two form factors,
\begin{eqnarray}
    \langle N(p',s') \vert A_\mu\vert N(p,s) \rangle = \bar{u}_N(p',s') 
    \bigg[\gamma_\mu G_A(Q^2) - \frac{Q_\mu}{2 m_N} G_P(Q^2)\bigg] \gamma_5 u_N(p,s),
    \label{Eq:DecompA}
\end{eqnarray}
where $G_A(Q^2)$ is the axial and $G_P(Q^2)$ the induced pseudoscalar
form factor. Here $Q^2=-q^2$ with $q=p'-p$ the momentum transfer and
$m_N$ the nucleon mass. The axial form factor at zero momentum
transfer gives the axial charge, $g_A \equiv G_A(0)$, while its slope
determines the axial radius,
\begin{equation}
    \langle r_A^2\rangle = - \frac{6}{ g_A} \frac{\partial G_A(Q^2)}{\partial Q^2} \bigg \vert_{Q^2 \rightarrow 0}.
\end{equation}

The nucleon matrix element of the pseudoscalar current $P=\bar{u}
\gamma_5 u - \bar{d} \gamma_5 d$ defines the pseudoscalar form factor,
\begin{equation}
    \langle N(p',s') \vert P \vert N(p,s) \rangle =  G_5(Q^2) \bar{u}_N(p',s')\gamma_5 u_N(p,s).
\end{equation}

These form factors are related through the partially conserved
axial-vector current (PCAC) relation,
\begin{equation}
    G_A(Q^2) - \frac{Q^2}{4 m_N^2} G_P(Q^2) = \frac{m_q}{m_N} G_5(Q^2),
\end{equation}
where $m_q$ is the light quark mass. Near the pion pole, assuming pion
pole dominance (PPD), the induced pseudoscalar form factor can be
expressed as,
\begin{equation}
    G_P(Q^2) = \frac{4 m_N^2}{Q^2+m_\pi^2} G_A(Q^2),
\end{equation}
where $m_\pi$ is the pion mass. Both PCAC and PPD relations will be
examined in the continuum limit of our lattice calculation. We also
compute the induced pseudoscalar coupling determined at the muon
capture point,
\begin{equation}
    g_P^* = \frac{m_\mu}{2 m_N} G_P(0.88\,m_\mu^2),
\end{equation}
where $m_\mu$ is the muon mass, as well as the pion-nucleon coupling
constant $g_{\pi NN}$ defined through the pseudoscalar form factor at
the pion pole.

\section{Lattice setup and statistics}

We describe the ensembles used in this work and detail the statistics
for computing correlation functions.

We use three \Nf{2}{1}{1} twisted mass fermion ensembles at the
physical point with lattice spacings spanning from 0.08 fm to 0.057
fm. The parameters of these ensembles are given in
Table~\ref{tab:ens}. The bare light quark mass parameter $\mu_l$ is
tuned to reproduce the isosymmetric pion mass $m_\pi=135$~MeV, while
the heavy quark parameters $\mu_s$ and $\mu_c$ are tuned via the ratio
of D-meson mass to decay constant and the ratio of the strange to
charm quark mass~\cite{Finkenrath:2022eon,Alexandrou:2018egz}.

\begin{table}[h]
    \caption{Parameters of the \Nf{2}{1}{1} ensembles analyzed in this
      work. We give the lattice volume, $\beta=6/g^2$ with $g$ the
      bare coupling constant, the lattice spacing $a$, the number of
      gauge configurations $N_{\rm conf}$, the pion mass $m_\pi$, and
      $m_\pi L$. Lattice spacings and pion masses are taken from
      Ref.~\cite{ExtendedTwistedMass:2022jpw}.}
    \label{tab:ens}    \centering
    \begin{tabular}{ccccccc}
    \hline\hline
       Ensemble  & $V/a^4$ & $\beta$ & $a$ [fm] & $N_{\rm conf}$ & $m_\pi$ [MeV]  & $m_\pi L$ \\
       \hline 
        \texttt{cB211.072.64} & $64^3 \times 128$ & 1.778 &  0.07957(13) & 750 & 140.2(2) & 3.62 \\
        \texttt{cC211.060.80} & $80^3 \times 160$ & 1.836 &  0.06821(13) & 400 & 136.7(2) & 3.78 \\
        \texttt{cD211.054.96} & $96^3 \times 192$ & 1.900 &  0.05692(12) & 500 & 140.8(2) & 3.90 \\
        \hline
    \end{tabular}
\end{table}

For each ensemble, we compute two- and three-point correlation
functions using multiple source positions per gauge configuration. For
two-point functions, we use 477, 650, and 480 source positions for the
three ensembles respectively. For three-point functions, we employ
seven to ten different sink-source time separations ranging from
approximately 0.5~fm to 1.5~fm. The number of source positions per
configuration is increased with the sink-source separation to maintain
approximately constant statistical errors, ranging from
$\mathcal{O}$(1) for the shortest separation to $\mathcal{O}$(100) for
the largest~\cite{Alexandrou:2023qbg}.

We neglect disconnected quark loop contributions in the present work
since in the twisted mass formulation these contributions to isovector
matrix elements are of order $a^2$ and thus vanish in the continuum
limit~\cite{Frezzotti:2003ni}.

The matrix elements are renormalized non-perturbatively using methods
based on Ward identities, which are fully non-perturbative and require
no gauge fixing~\cite{ExtendedTwistedMass:2022jpw}. This approach provides much
more accurate results on the renormalization constants compared to the
standard RI$^\prime$ scheme.

\section{Extraction of Form Factors}

The nucleon matrix elements are determined from two- and three-point correlation functions. The spectral decomposition of the two-point function is given by
\begin{align}
C(\Gamma_0,\vec{p},t_s) &= \sum_{i}^{N_{st}-1}c_i(\vec{p}) e^{-E_i(\vec{p}) t_s}
\label{Eq:Twp_tsf}
\end{align}
and the three-point function by
\begin{align}
C_{\mu}(\Gamma_k,\vec{q},t_s,t_{\rm ins}) &= \sum_{i,j}^{N_{st}-1}{\cal A}^{i,j}_{\mu}(\Gamma_k,\vec{q}) e^{-E_i(\vec{0})(t_s-t_{\rm ins})-E_j(\vec{q})t_{\rm ins}},
\label{Eq:Thrp_tsf}
\end{align}
with $t_s$ the sink time and $t_{\rm ins}$ the current insertion
time. The coefficients $c_i(\vec{p})$ are overlap terms of the
interpolating operator with the $i$-th state while ${\cal
  A}^{i,j}_\mu$ contain the matrix elements between states $i$ and
$j$. The desired ground-state matrix element is obtained by
$\frac{{\cal
    A}^{0,0}_\mu(\Gamma_k,\vec{q})}{\sqrt{c_0(\vec{0})c_0(\vec{q})}}$. The
coefficients $c_i(\vec{p})$ and ${\cal A}^{i,j}_\mu$ are determined by
simultaneous fits to two- and three-point functions at multiple
$t_s$. The sums of Eqs.~(\ref{Eq:Twp_tsf}) and (\ref{Eq:Thrp_tsf}) are
truncated at either $N_{st}=2$ (two-state fits) or $N_{st}=3$
(three-state fits). For the lowest non-zero momentum transfer, we
perform combined fits to matrix elements of the axial current
including both spatial and temporal components, to better constrain
the excited state energies in the fits. For each fit we vary i) the
minimum value of $t_s$ included in the fit range of the two-point
function ($t_{\rm 2pt,\,min}$), ii) the minimum value of $t_s$
included in the fit range of the three-point function ($t_{\rm
  3pt,\,min}$), iii) the number of insertion time slices kept near the
source ($t_{\rm ins,\,0}$) and sink ($t_{\rm ins,\,S}$), and iv) for
three-state fits, we include either only the terms ${\cal
  A}^{0,0}_\mu$, ${\cal A}^{1,0}_\mu$, ${\cal A}^{0,1}_\mu$, ${\cal
  A}^{1,1}_\mu$, ${\cal A}^{2,0}_\mu$, ${\cal A}^{0,2}_\mu$ ($N_O=6$),
or all terms of the three-state fit, i.e. including ${\cal
  A}^{2,1}_\mu$, ${\cal A}^{1,2}_\mu$, ${\cal A}^{2,2}_\mu$ ($N_O=9$).

\begin{figure}[t]
   \centering
   \includegraphics[width=0.8\linewidth]{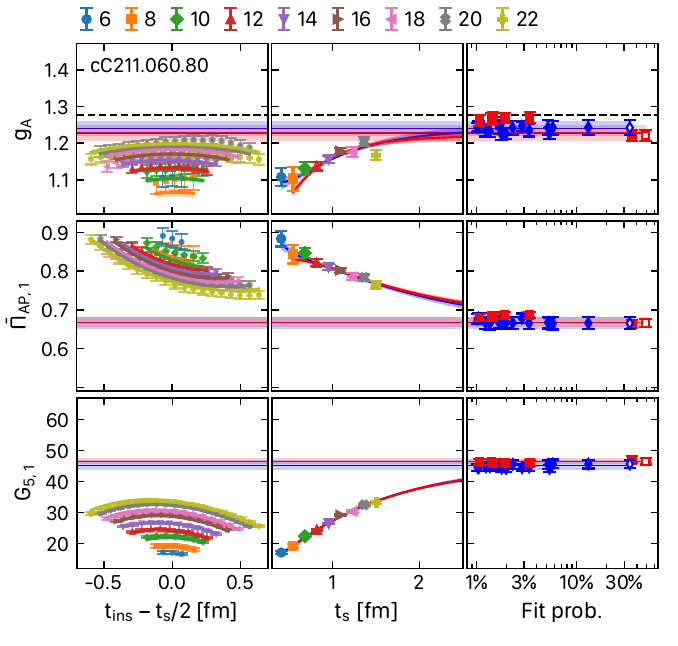}
   \caption{The ratio of Eq.~(\ref{Eq:ratio_new}) for the ensemble
     cC211.060.80 and for three cases, namely that yielding $g_A$
     (top), that yielding a linear combination of the axial and
     induced pseudoscalar form factor at the first non-zero momentum
     transfer (middle) and that yielding the pseudoscalar form factor
     at the first non-zero momentum transfer (bottom). In the left
     column we plot the ratio versus the insertion time, in the middle
     we plot for $t_{\rm ins}=t_s/2$ and in the right column we show
     the asymptotic value of the two- (blue) or three- (red) state
     fits versus the fit probability. In the left and middle columns,
     the curves correspond to the fit results which have the largest
     probability.}
   \label{fig:correlator_fits}
\end{figure}

The results from different fit ranges are combined using the Akaike
Information Criterion (AIC). To each fit $i$ we assign a weight
\begin{equation}
\log(w_i) = -\frac{\chi^2_i}{2} + N_{{\rm dof},i},
\end{equation}
where $N_{\rm dof} = N_{\rm data} - N_{\rm params}$ is the number of
degrees of freedom~\cite{Jay:2020jkz,Neil:2022joj}. The model averaged
value of an observable $\mathcal{O}$ is then given by
$\langle\mathcal{O}\rangle = \sum_i {\bar{\mathcal{O}}}_i p_i$ and its
error squared by $\sum_i (\sigma_i^2+\bar{\mathcal{O}}_i^2) p_i-
\langle O\rangle^2$, where $p_i = w_i/\sum_j w_j$ and
${\bar{\mathcal{O}}}_i$ and $\sigma_i$ are the central value and error
from fit $i$.

In Fig.~\ref{fig:correlator_fits} we provide an example analysis for
the intermediate lattice spacing, with results for all ensembles given
in Ref.~\cite{Alexandrou:2023qbg}. For visualization purposes, we
construct the ratio,
\begin{align}
R'_{\mu}(\Gamma_k;\vec{q};t_s,t_{\rm ins}) &= \frac{C_{\mu}(\Gamma_k,\vec{q};t_s,t_{\rm ins})}{\sqrt{C(\Gamma_0,\vec{0};t_s)C(\Gamma_0,\vec{q};t_s)}},
\label{Eq:ratio_new}
\end{align}
which at large time-separations $t_s-t_{\rm ins}\gg$ and $t_s\gg$
yields the ground-state matrix element.  Both two- and three-state
fits yield consistent results, with the most probable fits having
probabilities between 10\% and 50\%.

\section{Results for axial and pseudoscalar form factors}

We analyze the $Q^2$-dependence of the axial form factor using both
dipole and z-expansion parameterizations. The dipole Ansatz is given
by $G(Q^2) = \frac{g}{(1+\frac{Q^2}{m^2})^2}$, with the axial radius
given by $\langle r^2\rangle = 12/m^2$.

The z-expansion is given as $G(Q^2) = \sum_{k=0}^{k_{\rm max}} a_k\;
z^k(Q^2)$, where $z(Q^2) = \frac{\sqrt{t_{\rm cut} + Q^2} -
  \sqrt{t_{\rm cut}+t_0} }{ \sqrt{t_{\rm cut} + Q^2} + \sqrt{t_{\rm
      cut}+t_0} }$, with $t_{\rm cut}=(3m_\pi)^2$ and $t_0=0$. The
coefficients $a_k$ are constrained using Gaussian priors centered at
zero with width that falls like $1/k$ to ensure smooth convergence at
large $Q^2$.

\begin{figure}[t]
   \centering \includegraphics[width=1\linewidth]{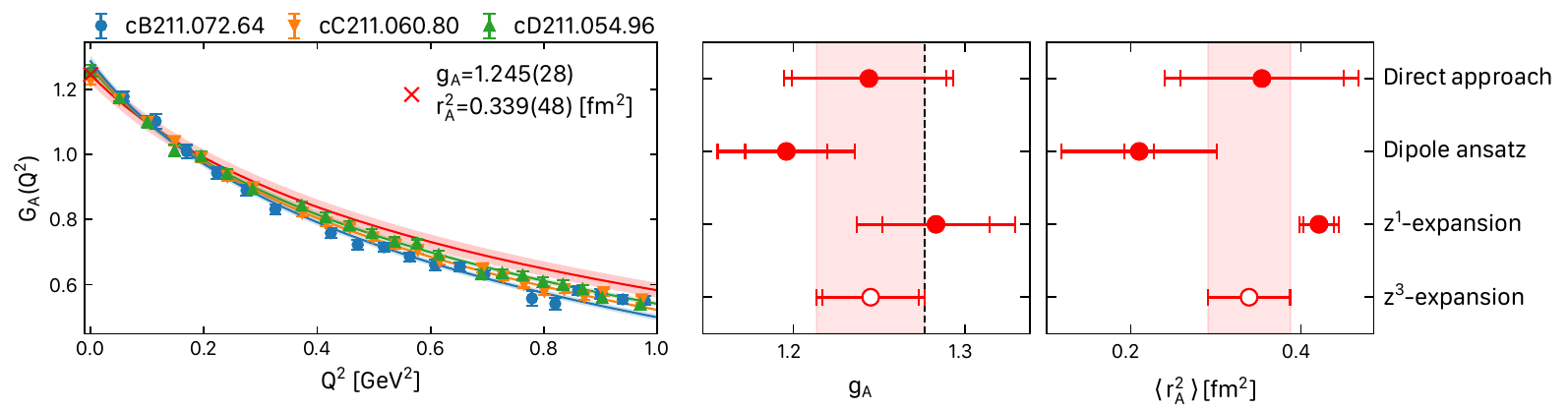}
   \caption{Left: The axial form factor $G_A(Q^2)$ determined from
     two-state fits (red band) with systematic uncertainty from the
     difference with three-state fits (yellow band). Results at finite
     lattice spacing shown in blue, orange and green bands. Right:
     Results for the axial charge $g_A$ and radius $\langle
     r_A^2\rangle$ obtained using different analysis approaches.}
   \label{fig:GA_final}
\end{figure}

For both parameterizations, we account for cut-off effects by allowing
a linear $a^2$ dependence in the parameters. The analysis is performed
in two ways, namely fitting the $Q^2$-dependence for each ensemble separately
followed by continuum extrapolation of the parameters (two-step), or
fitting all ensembles simultaneously (one-step). Both approaches yield
consistent results. We find convergence of the z-expansion at $k_{\rm
  max}=3$ and verify stability with respect to the prior width and
maximum $Q^2$ included in the fits. The analysis is performed
separately for the matrix elements extracted using two- and
three-state fits. The two-state fits allow analysis up to
$Q^2=1$~GeV$^2$, while three-state fits become unstable beyond
$Q^2\simeq 0.5$~GeV$^2$. The results from both analyses are consistent
within the range where three-state fits are stable. The resulting form
axial factor using a two-state fit analysis is shown in the left panel
of Fig.~\ref{fig:GA_final}.

Our final results use the z-expansion fits to the two-state data as
central values, with a systematic error taken as the difference
between the central values when using two- or three-state fits. The
values for $g_A$ and $\langle r_A^2\rangle$ using this approach are in
the right panel of Fig.~\ref{fig:GA_final}, where the direct
extraction is also shown, i.e. by computing the radius from $Q^2=0$
and the lowest non-zero $Q^2$. The consistency among these different
approaches, particularly between the direct approach and z-expansion
which uses data at all $Q^2$, demonstrates the robustness of our
analysis.

\begin{figure}[t]
   \centering
   \includegraphics[width=0.48\linewidth]{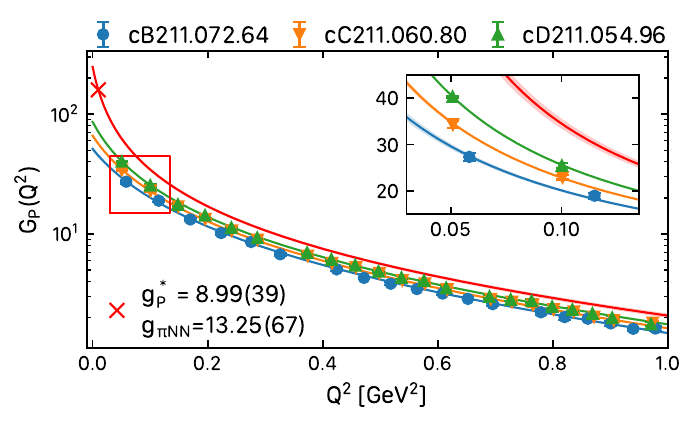}
   \includegraphics[width=0.48\linewidth]{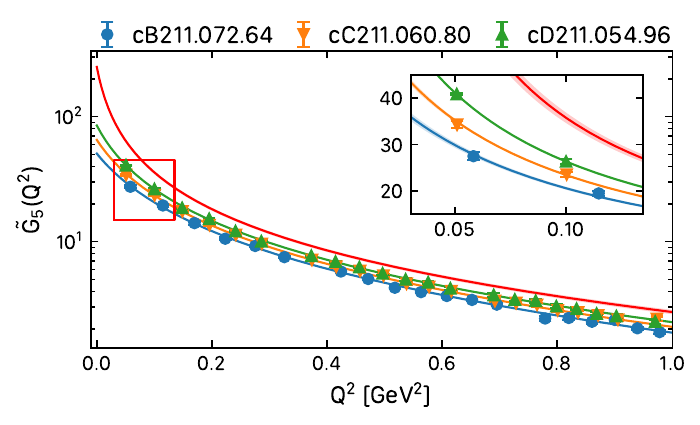}
   \caption{Left: The induced pseudoscalar form factor $G_P(Q^2)$ and
     right: the pseudoscalar form factor $\tilde{G}_5(Q^2)$ at finite
     lattice spacing (blue, orange and green bands) and in the
     continuum limit (red band). Results are obtained using two-state
     fits and z-expansion of order 3. The inner panels show the region
     near the pion pole enlarged.}
   \label{fig:GP_G5}
\end{figure}

The induced pseudoscalar $G_P(Q^2)$ and pseudoscalar $G_5(Q^2)$ form
factors exhibit a pion pole at $Q^2=-m_\pi^2$. For $G_5(Q^2)$ we
analyze the scaled quantity, $
   \tilde{G}_{5}(Q^2) = \frac{4m_N}{m_\pi^2} m_q G_5(Q^2),
$
where the combination $m_q G_5(Q^2)$ is scale-independent and
renormalizes with $Z_S/Z_P$. The scaling by $1/m_\pi^2$ accounts for
slight variations in the simulated pion masses, while $m_N$ makes the
combination dimensionless. We perform a combined fit of both form
factors using a third-order z-expansion after factoring out the pion
pole. Since the pion pole dominance relation is satisfied at the
continuum limit, we enforce the value of the pion-nucleon coupling
constant $g_{\pi NN}$ extracted from both form factors to be the
same. The resulting form factors are shown in Fig.~\ref{fig:GP_G5},
where the inner panels highlight the behavior near the pion pole.

\section{PCAC and pion pole dominance}
We examine the PCAC and PPD relations by constructing the ratios
\begin{align}
r_{\rm PCAC}(Q^2) &= \frac{\frac{m_q}{m_N} G_5(Q^2) + \frac{Q^2}{4 m_N^2} G_P(Q^2) }{G_A(Q^2)}\,\,\mathrm{and}\,\, &
r_{\rm PPD}(Q^2) &= \frac{m_\pi^2 + Q^2}{4 m_N^2}\frac{G_P(Q^2)}{ G_A(Q^2) }.
\end{align}
The PCAC relation requires $r_{\rm PCAC}=1$ for all $Q^2$, while PPD
predicts $r_{\rm PPD}=1$ near the pion pole.

\begin{figure}[!h]
   \centering
   \includegraphics[width=0.48\linewidth]{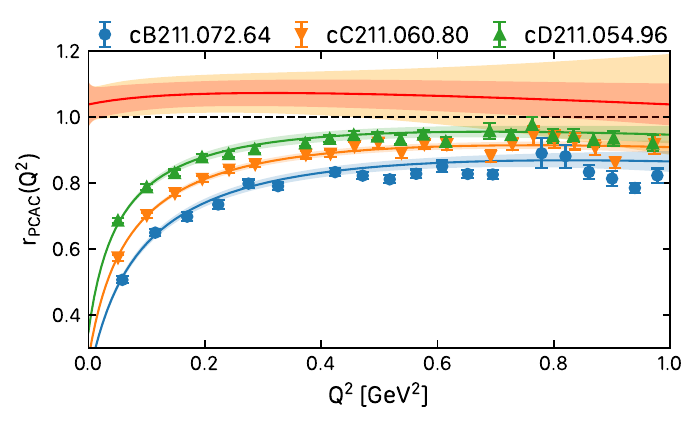}
   \includegraphics[width=0.48\linewidth]{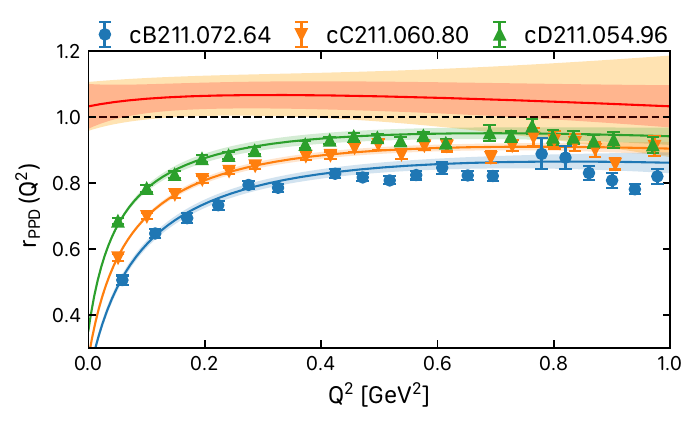}
   \caption{Left: Ratio testing the PCAC relation. Right: Ratio
     testing pion pole dominance. Results at finite lattice spacing
     shown in blue, orange and green bands and in the continuum limit
     with red band. The yellow band includes systematic uncertainties
     from excited states.}
   \label{fig:PCAC_PPD}
\end{figure}

In the twisted mass formulation at finite lattice spacing, we observe
sizable cut-off effects in both ratios. These arise from
$\mathcal{O}(a^2)$ effects in the pion mass that enter through the
pion pole in $G_P(Q^2)$ and $G_5(Q^2)$. The pion pole mass obtained
using valence Osterwalder-Seiler quarks in the mixed action
formulation shows significant cut-off effects, much larger than the
mass splitting between the unitary charged and neutral pion. However,
as shown in Fig.~\ref{fig:PCAC_PPD}, both relations are restored in
the continuum limit where the pion pole mass agrees with the physical
pion mass. Thus, at the continuum limit, our results reproduce these
fundamental relations that follow from chiral symmetry.

\section{Summary}

We have presented a lattice QCD calculation of the nucleon isovector
axial, induced pseudoscalar, and pseudoscalar form factors using three
ensembles simulated with quark masses that reproduce the physical pion
mass. The use of ensembles simulated solely at the physical point
eliminates systematic uncertainties from chiral extrapolation. The
three ensembles span lattice spacings from 0.08~fm to 0.057~fm with
approximately equal physical volumes, enabling a controlled continuum
extrapolation. Our analysis shows that both the PCAC relation and pion
pole dominance are satisfied in the continuum limit, despite sizable
cut-off effects at finite lattice spacing arising from the pion pole
in the twisted mass formulation.

Through a careful analysis of excited states and model averaging of
multiple fit variations, we obtain, in the continuum limit, the axial
charge ($g_A$), axial radius ($\langle r_A^2\rangle$), induced
pseudoscalar coupling ($g^*_P$), and pion-nucleon coupling constant
($g_{\pi NN}$),
\begin{equation}
\begin{aligned}
   g_A&=1.245(28)(14),                  & \langle r_A^2\rangle &=0.339(48)(06)~{\rm fm}^2,\\
   g^*_P&=8.99(39)(49),\,\,\textrm{and}\,\,\,                 & g_{\pi NN} &= 13.25(67)(69),
\end{aligned}
\end{equation}
where the first error is statistical from the model average of
two-state fits and the second is systematic from excited states. The
analysis presented here will be further improved by including a fourth
ensemble with lattice spacing $a=0.049$~fm and approximately same
physical volume as the three ensembles used here, with lattice volume
$112^3\times224$. This will provide an additional point in the
continuum extrapolation at an even finer lattice spacing, helping to
further constrain cut-off effects and improve the precision of our
final results. The analysis of this ensemble is ongoing, with first
results for charges presented in Ref.~\cite{charges:2024pos}.

\section*{Acknowledgments}
C.A., G.K., and G.S.  acknowledge partial support by the projects
3D-nucleon, NiceQuarks, and ``Lattice Studies of Strongly Coupled
Gauge Theories: Renormalization and Phase Transition''
(EXCELLENCE/0421/0043, EXCELLENCE/0421/0195, and EXCELLENCE/0421/0025)
co-financed by the European Regional Development Fund and the Republic
of Cyprus through the Research and Innovation Foundation as well as
AQTIVATE that received funding from the European Union’s research and
innovation program under the Marie Sklodowska-Curie Doctoral Networks
action, Grant Agreement No 101072344. C.A acknowledges support by the
University of Cyprus projects ``Nucleon-GPDs'' and
``PDFs-LQCD''. S.B. and J.F. are supported by the Inno4scale project,
which received funding from the European High-Performance Computing
Joint Undertaking (JU) under Grant Agreement No.~101118139. J.F. also
acknowledges support from the DFG research unit FOR5269 ``Future
methods for studying confined gluons in QCD'' and the Next Generation
Triggers project. M.C. acknowledges financial support from the
U.S. Department of Energy, Office of Nuclear Physics under Grant
No. DE-SC0020405, and the Grant No. DE-SC0025218. R.F. is supported by
the Italian Ministry of University and Research (MUR) under the grant
PNRR-M4C2-I1.1-PRIN 2022-PE2 Non-perturbative aspects of fundamental
interactions, in the Standard Model and beyond F53D23001480006 funded
by E.U.- NextGenerationEU. This work was supported by the Deutsche
Forschungsgemeinschaft (DFG, German Research Foundation) as part of
the CRC 1639 NuMeriQS – project no. 511713970. This work was supported
by grants from the Swiss National Supercomputing Centre (CSCS) under
projects with ids s702 and s1174. The authors gratefully acknowledge
the Gauss Centre for Supercomputing e.V. (www.gauss-centre.eu) for
funding this project by providing computing time through the John von
Neumann Institute for Computing (NIC) on the GCS Supercomputer
JUWELS-Booster at J\"ulich Supercomputing Centre (JSC). The authors
also acknowledge the Texas Advanced Computing Center (TACC) at The
University of Texas at Austin for providing HPC resources that have
contributed to the research results.

\bibliographystyle{JHEP} \bibliography{refs}

\end{document}